\documentclass{article}
\pdfoutput=1
 
\newif\ifIEEE
\IEEEfalse

\usepackage{arxiv}

\usepackage[utf8]{inputenc} 
\usepackage[T1]{fontenc}    
\usepackage{hyperref}       
\usepackage{url}            
\usepackage[nolist]{acronym}
\usepackage{graphicx}
\usepackage{subfig}
\usepackage{amsmath}
\usepackage{url}
\usepackage{booktabs}
\usepackage{multirow}
\usepackage{mathtools}
\usepackage{amssymb}

\makeatletter
\newcommand\footnoteref[1]{\protected@xdef\@thefnmark{\ref{#1}}\@footnotemark}
\makeatother

\begin{acronym}
    \acro{ACE}{Acoustic Characterization of Environments\acroextra{. A noisy reverberant speech corpus and IEEE challenge run by the SAP group at Imperial College}}
\acro{AIR}{Acoustic Impulse Response}
\acro{PCA}{Principal component analysis}
\acro{DNN}{Deep Neural Network}
\acro{DRR}{Direct-to-Reverberant Ratio}
\acro{FF}[FF]{Feed Forward}
\acro{FIR}{Finite Impulse Response\acroextra{. A filter whose output is a weighted sum of past input values and whose system function contains only zeros and no poles.}}
\acro{GAN}[GAN]{Generative Adversarial Network}
\acro{IIR}{Infinite Impulse Response\acroextra{. A filter whose output is a weighted sum of both past input and past output values and whose system function contains both poles and zeros.}}
\acro{RT}{Reverberation Time}
\acro{ReLU}[ReLU]{Rectified Linear Unit}
\acro{SED}[SED]{Sound Event Detection}
\acro{TOA}[ToA]{Time-of-Arrival}
\acro{WGN}{White Gaussian Noise}
\acro{GPU}[GPU]{Graphics Processing Unit}
\acro{TIMIT}{\acs{TI}-\acs{MIT}}
\acro{VAE}[VAE]{Variational Autoencoder}
\acro{CNN}[CNN]{Convolutional Neural Network}
\acro{RNN}{Recurrent Neural Network}

\end{acronym}

\title{Data Augmentation of Room Classifiers using Generative Adversarial Networks}

\author{
  Constantinos Papayiannis\\
  Dept. of Electrical and Electronic
Engineering\\
Imperial College London\\
London SW7 2AZ, U.K. \\
  \texttt{papayiannis@imperial.ac.uk} \\
   \And
 Christine Evers\\
  Dept. of Electrical and Electronic
Engineering\\
Imperial College London\\
London SW7 2AZ, U.K. \\
  \texttt{c.evers@imperial.ac.uk} \\
     \And
  Patrick A. Naylor\\
  Dept. of Electrical and Electronic
Engineering\\
Imperial College London\\
London SW7 2AZ, U.K. \\
  \texttt{p.naylor@imperial.ac.uk}
}

\begin{document}
\maketitle

\begin{abstract}
The classification of acoustic environments allows for machines to better understand the auditory world around them. The use of deep learning in order to teach machines to discriminate between  different rooms is a new area of research. Similarly to other  learning tasks, this task suffers from the high-dimensionality and the limited availability of training  data. Data augmentation methods have proven useful in addressing this issue in the tasks of sound event detection and scene classification. This paper proposes a method for data augmentation for the task of room classification from reverberant speech. Generative Adversarial Networks (GANs) are trained that generate artificial data as if they were measured in real rooms. This provides additional training examples to the classifiers without the need for any additional data collection, which is  time-consuming and often impractical. A representation of acoustic environments is proposed, which is used to train the GANs. The representation is based on a sparse model for the early reflections, a stochastic model for the reverberant tail and a mixing mechanism between the two. In the experiments shown, the proposed data augmentation method increases the test accuracy of a CNN-RNN room classifier from 89.4\% to 95.5\%.

\end{abstract}


\section{Introduction}

The reverberation effect is present in all real life enclosures and provides a listener with cues that relate to properties of the room. These cues are  the cumulative result of many  acoustic reflections that   human listeners use to infer properties of the room \cite{Dokmanic2011}. Similarly, models can be learned by machines that enable them to infer properties of their auditory environments \cite{Papayiannis2017}. Training data available for  learning properties of the reverberation effect   is often in the form of \acp{AIR} as \ac{FIR} filters, which are measured in real rooms \cite{Farina2000}. These \acp{AIR} are high-dimensional, consisting  typically of thousands of coefficients, and they are small in number as their measurement is time-consuming \cite{Farina2000} and often impractical. This  limits the training of  classifiers based on \acp{DNN} \cite{Papayiannis2018a}. The motivation of this paper is to address this issue for the task of room classification, where a machine is trained to predict the room where a speech recording was made in. This finds applications in a smart-home \cite{Papayiannis2018a}, providing machines with understanding about the location of the user in the home, and also in forensics \cite{Moore2014}.

This paper presents a novel method for data augmentation for room classification from reverberant speech.  The data augmentation  method   starts from measured \acp{AIR} and  uses \acp{GAN} to generate additional artificial ones. To do so, one \ac{GAN} is trained for each of the rooms considered in the classification and it is then used to generate many artificial \acp{AIR}. This is an alternative to the process of measuring many more \acp{AIR}, by moving the source and receiver at various positions in the same real room.  Repeating the process for a number of rooms expands the available dataset, without the need for any additional data collection. A challenge to overcome during  training is related to the motivation for this work, which is the high-dimensionality of  \acp{AIR}. This is overcome  by using a proposed low-dimensional representation for acoustic environments. The representation describes sparse early reflection using the  parameters estimated in \cite{Papayiannis2017a} and uses established acoustic parameters to represent the late reverberation. Creating a low-dimensional representation  also allows for the evaluation of the generated responses and their distribution across a set of parameters relevant to the task. Evaluating the  data generated by \acp{GAN} is typically not straightforward, which a drawback  in their use \cite{Salimans2016}. In this work, the generated samples consist of a  small and semantically meaningful parameter set, which allows for easier evaluation of the results.   In the experiments shown, the  data augmentation method improves the accuracy of the  \acs{CNN}-\acs{RNN} room classifier proposed in \cite{Papayiannis2018a}. To illustrate the effectiveness of using the proposed low-dimensional representation of \acp{AIR}, the experiments shown compare it with the use of the raw \ac{FIR} taps. The \acp{AIR} generated by the \acp{GAN}  find uses beyond the data augmentation of classifiers, such as artificial reverberation \cite{Valimaki2012}.

The remainder of this paper is organised as follows: Section \ref{section_c8_data_augmentation} discusses  data augmentation for classification and Section \ref{section_c10_est_gen} presents the proposed method for generating artificial \acp{AIR} for room classification training. The experiments in Section \ref{chapter14} present the results of the proposed method. Finally, Section \ref{chapter15} provides a discussion of the results of the experiments and a conclusion.

\section{Data augmentation for classifier training}
\label{section_c8_data_augmentation}

The supervised training of classifiers relies on the collection of labelled data, serving as the examples the classifier learns from. In \cite{Papayiannis2018a}, \acp{DNN} were presented with a set of \acp{AIR} in order to learn to discriminate between different rooms. In realistic scenarios, it is impossible for the training data to  cover every point in the corresponding physical space. This  means that unseen data will be presented to the classifier during inference when a speech recording is made at a source and receiver position not part of the original data collection.   Substantially expanding the training data set  for   room classification and other tasks such as  \ac{SED}  is in general challenging \cite{Salamon2016}.  However, methods exist for   increasing the amount of available training   data  without the need for additional data collection. This process is referred to as data augmentation and uses the available training data to provide the classifier with  class invariant transformations of already seen examples \cite{Goodfellow2016}. This aims to increase the accuracy  of the classifier during inference by improving its generalisation.

The concept of data augmentation has been studied extensively in the literature in order to improve the accuracy of classifying audio signals. A very simple yet representative example of this concept is discussed in \cite{Takahashi2016}, which describes the task of detecting bird singing. In this example,  any segment of audio containing bird singing would be a positive sample. Still, any mixture of two, or more, positive samples would also be positive. This simple mechanism of overlapping audio segments allows for the expansion of the amount of available training data    with a simple overlap of two existing recordings. This \textit{manual} method for data augmentation does not involve a statistical model  but a simple logical reasoning and human understanding of the task. Other such methods  discussed in the literature include time-stretching of segments \cite{Parascandolo2016}, pitch shifting \cite{Schluter2015} and dynamic range compression \cite{Salamon2016}. A data augmentation method for audio data, which does not rely on such \textit{manual} processes is proposed in this paper. The focus is  the classification of reverberant rooms and the method relies on generative models, able to generate  additional artificial \acp{AIR}. 

The next Section discusses how \acp{DNN} are used to estimate generative models for different categories of reverberant rooms. 

\section{Generative model estimation for reverberant rooms}
\label{section_c10_est_gen}

The above Sections have discussed the motivation for estimating   generative models that allow the generation of artificial \acp{AIR} corresponding to real reverberant rooms and how this can improve the generalisation of room classifiers. 

\subsection{Estimation method}

A generative model represents the joint probability $P(\textbf{x},\textbf{y})$, which is in contrast to  classification \acp{DNN} that estimate the posterior $P(\textbf{x} \vert\textbf{y})$. Recent advancements in deep learning led to the  proposal of  alternatives to the traditional method for the   estimation of parametric model distributions. The two dominant methods in the modern literature  are \acp{GAN} and \acp{VAE}. Both follow a similar formulation that uses back-propagation to train network layers, which are able to estimate the generative model by filtering  noise  drawn from a known prior. In the literature review conducted for this work, \acp{GAN} have shown to be widely adopted in the field of audio processing  across different tasks such as  \ac{SED} \cite{Mun2017}, speech recognition \cite{Sriram2017}, speech enhancement \cite{Donahue2017} and dereverberation \cite{Li2018,Wang2018}. Furthermore, variants of the original \ac{GAN} in \cite{Goodfellow2014} exist, which can be adapted in the future to lead to more exciting applications of the method proposed in this work, such as  Conditional \acp{GAN} \cite{Mirza2014}, Dual\acp{GAN} \cite{Yi2017} and many others. \acp{GAN} are therefore chosen as the estimation mechanism for the generative models in this paper.

The rest of this Section discusses how the networks are trained to learn properties of the reverberation effect. Two options are explored, the first one uses the raw \ac{FIR} taps of \acp{AIR} to train the \acp{GAN}. The second uses  a proposed low-dimensional representation of the  \acp{AIR} to train the \acp{GAN}.

\subsection{\acs{GAN} training}
\label{section_c8_gan_training}

\acp{GAN} are composed of two networks that are posed as adversaries. The two networks play the roles of the generator and discriminator  \cite{Goodfellow2014}. The task of the discriminator $D(\mathbf{y};\theta_d)$ is to judge whether a given sample comes from the original data distribution or not. The task of the generator $G(\mathbf{y} \vert \mathbf{z} ;\theta_g)$, on the other hand, is to \textit{fool} the discriminator into thinking that data samples it produces are  originating from the original data distribution. $\mathbf{z}$ represents a random vector variable as $\mathbf{z} \sim N(\mathbf{0},\mathbf{I})$.

\ifIEEE
\newcommand{\thisscale}{0.4}
\else
\newcommand{\thisscale}{0.5}
\fi

\begin{figure}[t]
	\centering
    \subfloat[Generator Network]{
                \includegraphics[scale=\thisscale]{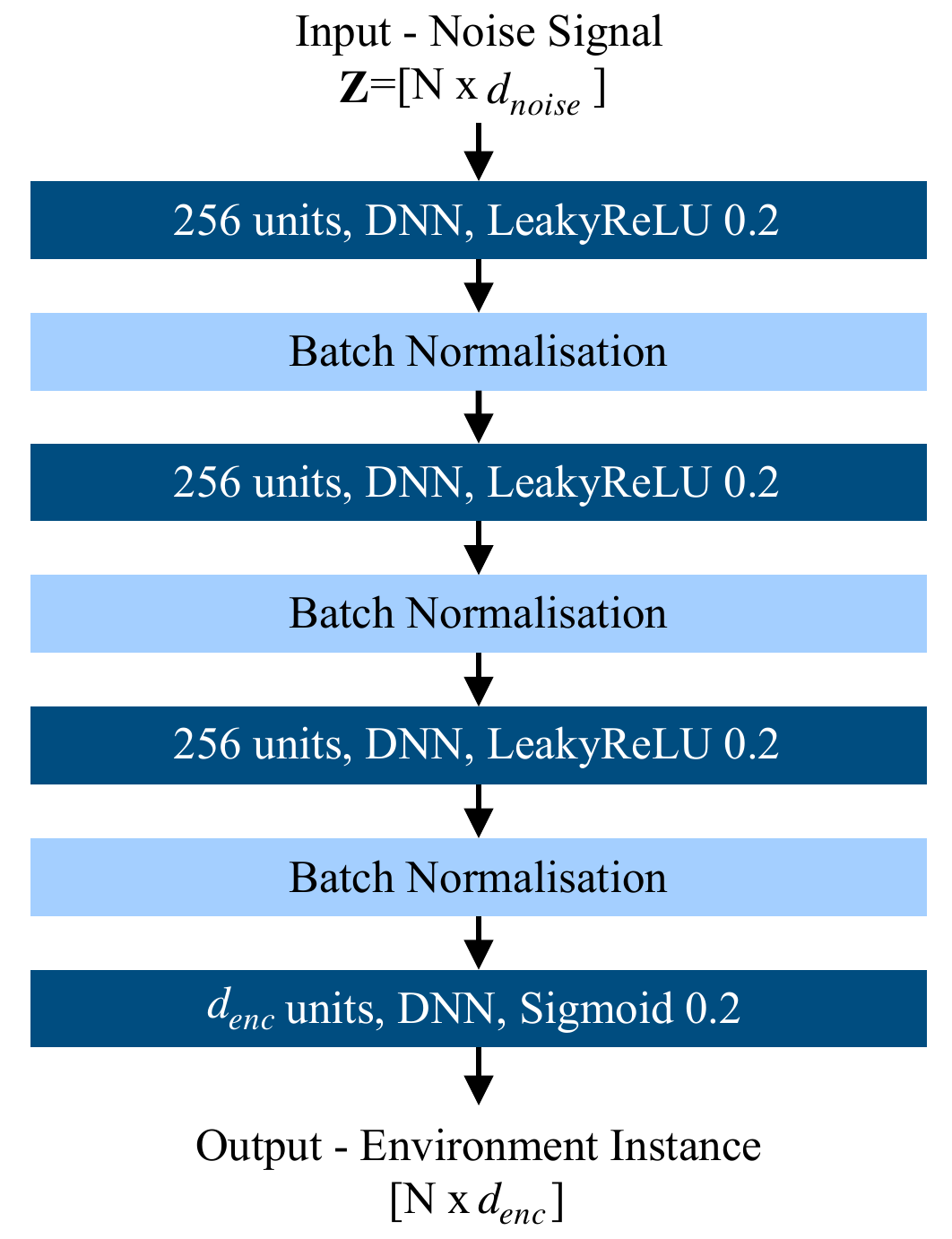}
                \label{figure_c10_generator}
        }
    \subfloat[Discriminator Network]{
                \includegraphics[scale=\thisscale]{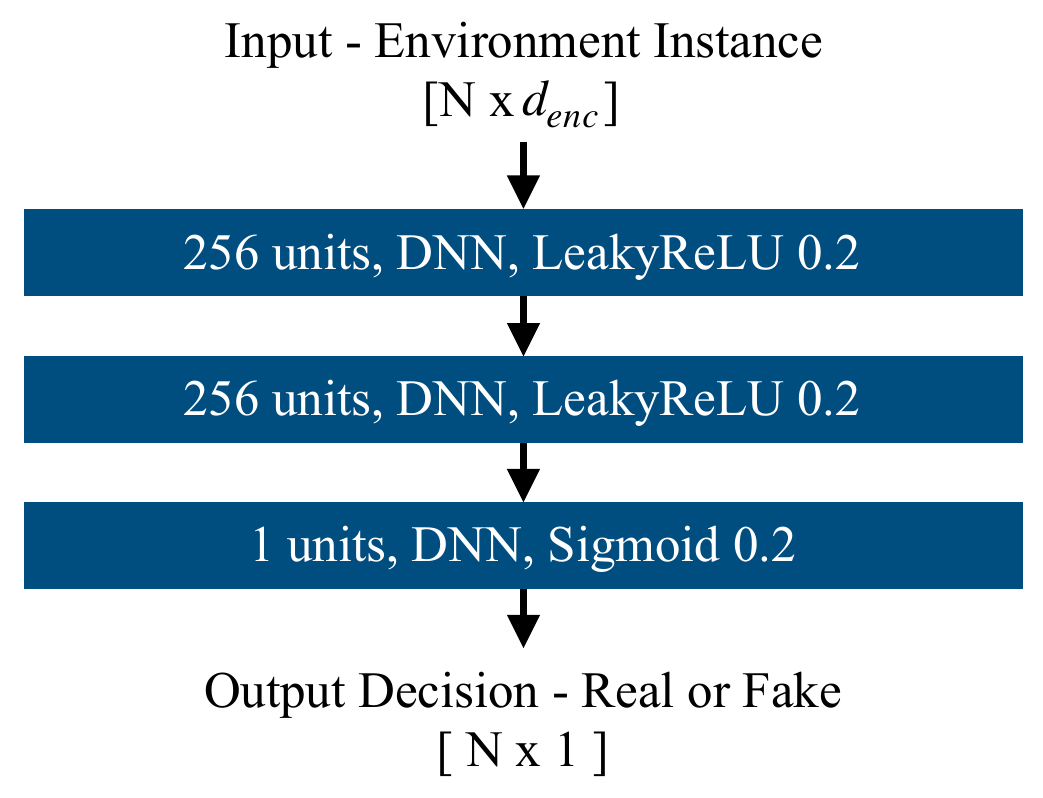} 
                \label{figure_c10_discriminator}
        }
	\caption{Generator and discriminator networks of \acp{GAN} used in this work. A \ac{GAN} is used to learn how to generate data as if they were measured in a specific room. Training one such \ac{GAN} for each room in a dataset allows for the generation of artificial responses from all rooms, expanding the available training data.}
	\label{figure_c10_generator_discriminator}
\end{figure}

The networks used in this work as the generator and discriminator  of  \acp{GAN} are shown in Figure \ref{figure_c10_generator_discriminator}. A simple \ac{DNN} architecture is used, composed only from \ac{FF} layers. More complex architectures can be constructed that include convolutional and recurrent layers. The investigation of the benefits of using other types of layers, as well as techniques such as dropout, is reserved for future work. Given the small size of the network, Leaky\acs{ReLU} activations \cite{Goodfellow2016} are used  instead of standard \acp{ReLU}, to counter issues that result   from  learning from gradients when the \ac{ReLU} activations are 0. Batch normalisation is used in order to improve the training, as proposed in  \cite{Ioffe2015}, by normalising the mean and standard deviation of activation. The networks are trained using back-propagation with Adam \cite{Kingma2014} as the optimizer. Inputs to the network are scaled to be within the range 0 and 1. The outputs are denormalised to restore the original scales. One normaliser-denormaliser pair is designed for each input-output neuron pair.  The training is run for a total of 6000 epochs. For the reasons relating to stability discussed in \cite{Arjovsky2017a}, additive \ac{WGN} is added to the inputs of the discriminator with $\sigma = 0.1$.

As part of the data augmentation process proposed in this paper, a \ac{GAN} is trained  given a set of training \acp{AIR} measured in a specific room and learns how to create new \acp{AIR} as if they were measured in the same room. Therefore, one \ac{GAN} is trained for each room considered. The \acp{GAN} discriminator uses \acp{AIR} to learn what is a \textit{real} \ac{AIR} and the generator uses them to learn how to imitate them and create \textit{fake} ones.     The rest of this Section investigates two choices for the way that \acp{AIR} are presented and outputted by the network.

\subsection{\acsp{GAN} using the \acs{FIR} filter taps of \acs{AIR}}
\label{section_c10_gan_air_domain}

In order to enable the discriminator and generator of \acp{GAN} to respectively idenify and generate realistic data, measured \acp{AIR} are presented to them during training. The simplest way to present \acp{AIR} to the networks during training is  using the  taps of  \ac{FIR} filters. The taps represent the sound pressure at the position of a receiver placed in the room, with the room excited by a source placed within its boundaries. This is the raw format in which \acp{AIR} are typically measured  \cite{Farina2000} and distributed \cite{Eaton2015d} in. An \ac{AIR} measurement as an \ac{FIR} filter is represented as a column vector whose elements are the taps of the filter. For $M$ \acp{AIR} measured in real rooms, their representation is  $\mathbf{h}_m$, where $m \in \{1,2,\dots,M\}$. The ideal discriminator's behaviour, in this case, is   therefore ${D(\mathbf{h}_m) = 1~\forall~m}$. The ideal generator's behaviour is $D(G(\mathbf{z})) = 1 $, where $\mathbf{z} \sim N(\mathbf{0},\mathbf{I})$. 

\subsection{\acsp{GAN} using a low-dimensional representation}
\label{section_c10_gan_expert_domain}

An alternative to processing and generating  \acp{AIR}  as \ac{FIR} filters   is proposed in this paper. Describing the \ac{AIR} as an \ac{FIR} filer is a typical choice but leads to a sequence of thousands of taps to be processed by algorithms. An alternative low-dimensional representation of the acoustic environment leads to fewer parameters to be processed, potentially improving the efficiency and effectiveness of training.  The proposed representation combines the early reflection parameters, estimated using the method of \cite{Papayiannis2017a}, with a set of established parameters for describing late reverberation to represent the training \acp{AIR}. With the training \acp{AIR} represented in this space, the trained \acp{GAN} will learn to model the distribution of each of the parameters, instead of how to model individually the thousands of \ac{FIR} taps. Furthermore, generated \acp{AIR} will be in this low-dimensional space, reducing the complexity of the generator and discriminator. \ac{FIR} filters that represent the  generated \acp{AIR} can be constructed from this low-dimensional representation.

The rest of this Section describes how \acp{AIR} as \ac{FIR} filters  are used to estimate the  parameters of the proposed low-dimensional representation. Also, the inverse process is described, which uses the low-dimensional representation to construct \ac{FIR} filters.

\subsubsection{Proposed low-dimensional representation}
\label{section_low_dim_rep}

The aim of the data augmentation method proposed in this paper is  to improve the generalisation of room classifiers during inference. Additional   examples  are generated that are  class invariant transformations of the available training data.  The task of room classification is to identify a known room at unknown  source and receiver  positions. With the transformation being class invariant, the available training \acp{AIR} from a room will be used to artificially generate  \acp{AIR} at new  source and receiver positions from the same room.  As highlighted in \cite{Papayiannis2017a}, early reflections have a strong and distinct structure in \acp{AIR}, which is highly related to these  positions. Therefore, manipulating  the structure  of early reflections  corresponds to a manipulation of these  positions.   Using a parametric representation of the early reflections, this paper uses \acp{GAN} to learn the distribution of   parameters of reflections from data measured within the room that enables the generation of many artificial responses. These responses will be generated as if they were measured in the same room but with the source and receiver positions changing.

A method was proposed in \cite{Papayiannis2017a}  for estimating the \acp{TOA} $\boldsymbol{\kappa}$ and scales $\boldsymbol{\beta}$ of $D$ early reflections in an \ac{AIR} and the excitation $h_e(n)$ that was used to measure it.  Modelling early reflections in this manner exploits knowledge about their sparse nature and enables the reconstruction of  the original \ac{FIR} taps as
\begin{equation}
h_r(n) = \sum_{i=1}^{D} {\beta_i} \left[  h_e(n){\ast}\frac{\sin\left[\pi(n-k_i)\right]}{\pi(n-k_i)} \right].
\label{eq_c10_reflections}
\end{equation}
The model can be complemented to represent the entire \ac{AIR}. In \cite{Lindau2012}, it was shown that  after a mixing point, defined as the tap with index $n_m$, replacing the original taps of \acp{AIR} by their reconstruction from a   stochastic model led to perceptually indistinguishable results. To construct a low-dimensional representation for the entire \ac{AIR}, this paper combines  (\ref{eq_c10_reflections}) with  a stochastic model for late reverberation that is based on Polack's model \cite{Naylor2010b}. Polack's model is described as 
\begin{equation}
	h_\text{Polack}(n) =  \nu(n) \exp^{-\frac{-3 n T_s \log 10}{T_{60} } },
	\label{polack_model}
\end{equation}
where $T_s$ the sampling period, $T_{60}$ is the reverberation time of the room and $\nu(n)\sim N(0,1)$. The expression shows a \ac{WGN} process, enveloped by an exponential decay term. This resembles the decaying sound level in a reverberant room after diffusion \cite{Kuttruff2009a}, which is  commonly referred to as the reverberant tail \cite{Naylor2010b} of an \ac{AIR}.  The stochastic model that is used in this work is based on (\ref{polack_model}) and includes terms that account for the difference in the decay rates of sound energy at different frequencies. This is done by filtering the tail signal by an \ac{IIR} filter, with numerator and denominator coefficients $\mathbf{b}$ and $\mathbf{a}$ respectively, estimated from the  tail of the original \ac{AIR}. In \cite{Haneda1994}, the   zeros and poles at  receiver positions in reverberant rooms were analysed and it was shown that both represent properties of the environment. Poles describe properties of the enclosure, whereas zeros vary with position. An \ac{IIR} filter with  $P$ zeros and $R$ poles is therefore designed to convey these properties to the reverberant tail.  The \ac{IIR} coefficients are applied to Polack's model to give the filtered reverberant tail 
\begin{equation}
	h_\text{tail}(n) =  \sum_{i=0}^P b_i  h_\text{Polack}(n-i) -  \sum_{j=1}^R a_j  h_\text{tail}(n-j).
\end{equation}
 A cross-fading mechanism is used to avoid abrupt discontinuities at  sample $n_m$, where the early reflections are mixed with the stochastic model. The mechanism is applied to the tail  to allow it to fade-in to a maximum of unity at $n_m$ and have symmetric values around it, giving the late reverberation model
\begin{equation}
	h_\text{late}(n) = 
\begin{dcases*}
0, & if $n<k_d$ \\
\frac{h_\text{tail}(2 n_m - n +k_d)}{{h_\text{tail}(0)}}, & if $k_d \leq n < n_m$ \\
\frac{h_\text{tail}(n -n_m -k_d)}{{h_\text{tail}(0)}}, & otherwise.
\end{dcases*}
\label{eq_c5_final_tail}
\end{equation}
Early reflections are described by  (\ref{eq_c10_reflections}) and the direct sound by 
\begin{equation}
	h_d(n)={\hat{\beta}_d}\hat{h}_e(n){\ast}\frac{\sin \left[ \pi(n-\hat{k}_d) \right]}{\pi(n-\hat{k}_d)},
\end{equation}
with $k_d$ and $\beta_d$ the \ac{TOA} and scale of the direct sound. The early reflections and the late reverberation model are scaled according to the \ac{DRR} \cite{Naylor2010b} values $\eta_1$ and $\eta_2$. They measure the energy ratio between the direct sound and the early reflections and the direct sound and the reverberant tail in the original \ac{AIR} and impose the same ratios on its reconstruction. The complete model, reconstructing taps of the \ac{FIR} filter representation of the \ac{AIR} is given by
\begin{equation}
	\hat{h}(n) = h_d(n) + \sqrt{\frac{\sum h_d^2(n)}{\eta_1  \sum h_r^2(n)}} h_r(n) + \sqrt{\frac{\sum h_d^2(n)}{\eta_2  \sum {h}^2_\text{late}(n)} }{h}_\text{late}(n).
\label{equation_chaplate_final}
\end{equation}

This paper proposes the use of   a  low-dimensional representation of \acp{AIR} to train \acp{GAN}, which is based on the formulation presented above. The parameters forming the representation are estimated from the  original \ac{AIR} taps and are able to reconstruct them. The parameters are the following:
\begin{itemize}
\itemsep0em
	\item [$\boldsymbol{\kappa}$, $\boldsymbol{\beta}$:] The \acp{TOA} and scales of the $D$ reflections up to 24~ms, estimated as proposed  in \cite{Papayiannis2017a}.
	\item [$\eta_1$, $\eta_2$:]  The two \ac{DRR} values discussed above, measured from the original \ac{AIR}.
	\item [$\mathbf{a}$, $\mathbf{b}$:] The coefficients of the \ac{IIR} filter for the reverberant tail, estimated using Prony's method \cite{Parks1987} from the tail of the original \ac{AIR}.
	\item [$T_{60}$:] The \ac{RT} of the environment, estimated using \cite{Karjalainen2002}.
\end{itemize}
All  the above vectors are defined as column vectors. One column vector with fixed-length is used to represent each \ac{AIR}.   It is created using the above parameters and it is used for training the \acp{GAN}. This column vector has a fixed-length   and for  \ac{AIR} $\mathbf{h}_m$ it is expressed as
\begin{equation}
	\tilde{\mathbf{h}}_m =  \left[ ~T_{60},  \eta_1, \eta_2, \mathbf{a}^T, \mathbf{b}^T,\mathbf{0}_{(D_{\text{max}} - D)},\boldsymbol{\kappa}^T,\mathbf{0}_{(D_{\text{max}} - D)},\boldsymbol{\beta}^T ~\right]^T.
	\label{eq_c10_encoding}
\end{equation}
The row vector $\mathbf{0}_{(D_{\text{max}} - D)}$  is used to account for the fact that the number of early reflections detected in each \ac{AIR} varies. 

\ifIEEE
\newcommand{\thisenva}{figure*}
\newcommand{\thisscaleb}{0.28}
\else
\newcommand{\thisenva}{figure}
\newcommand{\thisscaleb}{0.22}
\fi

\begin{\thisenva}[t]
	\centering
	\includegraphics[scale=\thisscaleb]{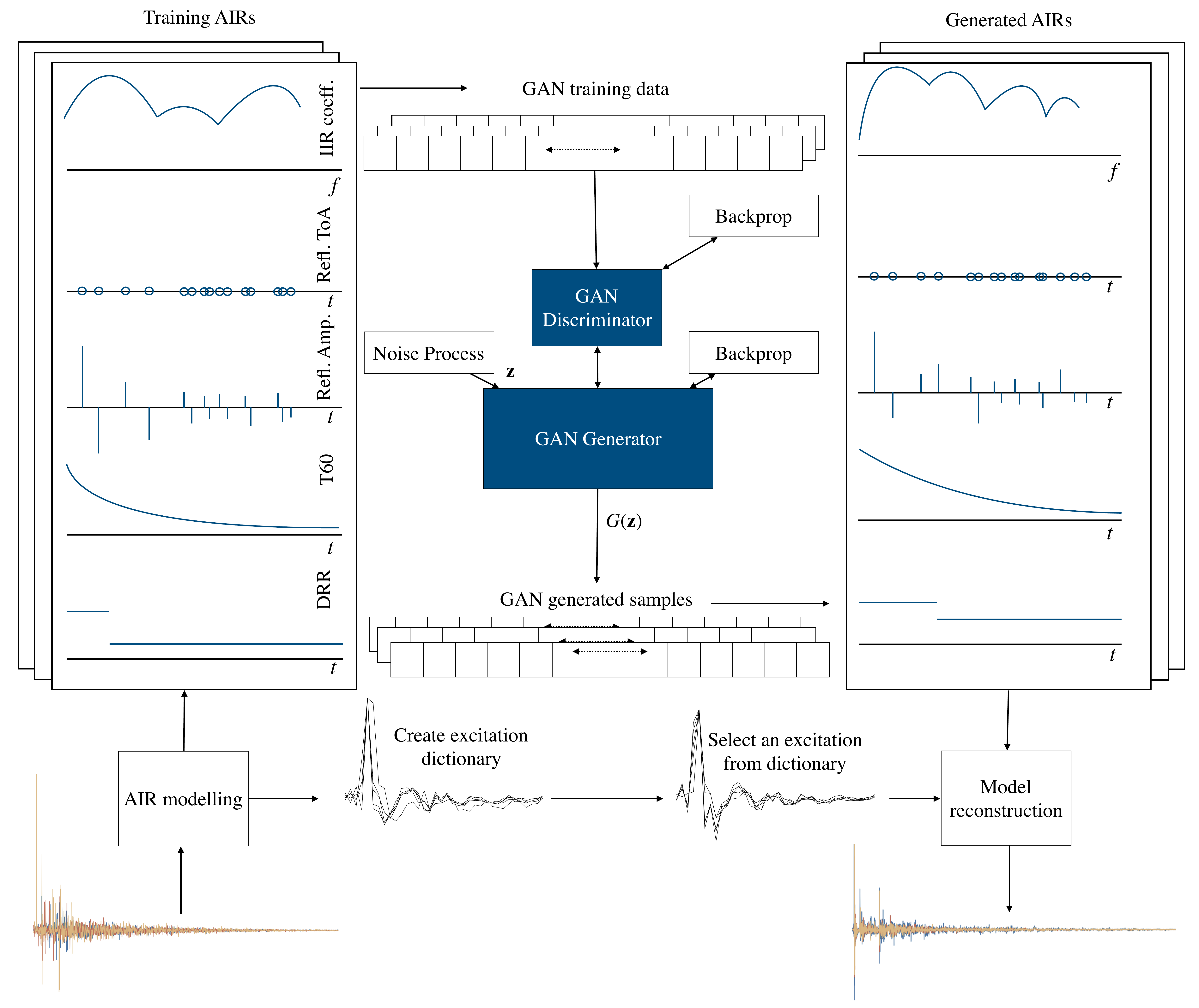}
	\caption{Training \acsp{GAN} to generate \acp{AIR} using the proposed low-dimensional representation of the training \acp{AIR}. The representation  consists of parameters describing the early reflections and a description of the effect of late reverberation. The parameters of the generated \acp{AIR}  then  construct the taps of \ac{FIR} filters, which are used as additional training data for room classifier  training.  }
	\label{figure_chapter10_gan_training_ac_env}
\end{\thisenva}

The excitation   $h_e(n)$ in (\ref{eq_c10_reflections}) accounts for the non-idealities in the method and equipment that was used to measure the \ac{AIR}. Including this excitation to in the final reconstruction of the \ac{AIR} is important. Skipping this step and using $h_e(n)=1$ as the excitation represents the use of a source and receiver system with linear response and infinite bandwidth, which is unrealistic. In this work, a bank of excitations is estimated from the training data. They are \ac{PCA} reconstructions of direct-path windows of measured \ac{AIR}. Using a number of principal components which explain 95\% of the variance in the samples of the windows gives realistic excitations and avoids the inclusion of overlapping reflections. Generated \acp{AIR} from the \acp{GAN} are then reconstructed using a randomly chosen excitation from the bank.

The \ac{IIR} filter with coefficients $\mathbf{a}$ and $\mathbf{b}$  involves $P$ zeros and $R$ poles. It conveys information about resonances and the spectrum of reverberant speech recorded in the room \cite{Haneda1994}.  Any poles that are part of the filter that are outside of the unit circle will  lead to an unstable system. To prevent this, a zero-pole analysis is performed on generated values for $\mathbf{a}$ and $\mathbf{b}$ and any poles outside of the unit circle are removed. With the focus being on low-dimensional representations, the small values of $P=R=5$ are chosen.  The order  selection for \ac{IIR} models for \acp{AIR} is discussed in \cite{Evers2011}.

The overall process that is described in this Section is summarised in Figure \ref{figure_chapter10_gan_training_ac_env}. The following Sections present experiments that evaluate the effectiveness of generating \acp{AIR} for data augmentation for the training of room classifiers. The experiments  will first present and analyse the  generated \acp{AIR}. Their usefulness will be measured later in   terms of the gains in accuracy provided for the task of room classification.

\section{Experiments}
\label{chapter14}

This paper proposes a novel method for data augmentation for the training of \ac{DNN} room classifiers. The method relies on the training of \acp{GAN}, which are used to generate artificial  \acp{AIR} that  increase the  training data available for the classifiers. The experiments described in this Section  illustrate the data generated by the \acp{GAN} and room classification experiments  evaluate their efficacy in data augmentation.   The proposed method relies on a low-dimensional representation of \acp{AIR} and to highlight its usefulness  it is compared to the use of the raw taps of \ac{FIR} filters for the training of \acp{GAN}.

The  dataset used to train the  \acp{GAN}  is a set of  \acp{AIR} provided with the \ac{ACE} challenge database \cite{Eaton2015d}. A total of 658 responses are used to train the \acp{GAN}, split evenly across 7 rooms. The  \acp{AIR} are padded to a duration of 2.1~s, the length of the longest \ac{AIR} in the training dataset. All data is downsampled to 16~kHz.

\subsection{\acs{AIR} generation}

\newcommand{\mpwidth}{0.4}

\begin{table}[t!]
\centering
\subfloat[\ac{FIR} taps.]{
\centering
\begin{tabular}{@{}lrrr@{}}
\toprule
Layer & \begin{tabular}[r]{@{}r@{}}Param.\\ count\end{tabular}  & \begin{tabular}[r]{@{}r@{}}Input\\ Dim.\end{tabular} \\  \midrule
\multicolumn{3}{c}{Generator Network}\\  \midrule
FF & 5,376 & 20 \\
Batch N. & 1,024 & 256 \\
FF & 65,792 & 256 \\
Batch N. & 1,024 & 256 \\
FF & 65,792 & 256  \\
Batch N. & 1,024 & 256 \\
FF & 8,544,736 & 256 \\  \midrule
\multicolumn{3}{c}{Discriminator Network}\\  \midrule
FF & 8,511,744  & 33,248  \\
FF & 65,792  & 256  \\
FF & 257  & 256  \\ \bottomrule
Total & 17,262,561 & 1 
\end{tabular}
\label{table_c10_params_e2e}
}~~~
\subfloat[Proposed low-dimensional representation.]{
    \centering
\begin{tabular}{@{}lrrr@{}}
\toprule
Layer & \begin{tabular}[r]{@{}r@{}}Param.\\ count\end{tabular}  & \begin{tabular}[r]{@{}r@{}}Input\\ Dim.\end{tabular}\\  \midrule
\multicolumn{3}{c}{Generator Network}\\  \midrule
FF & 5,376  & 20  \\
Batch N. & 1,024  & 256 \\
FF & 65,792  & 256 \\
Batch N. & 1,024  & 256 \\
FF & 65,792  & 256 \\
Batch N. & 1,024  & 256 \\
FF & 43,690  & 256 \\ \midrule
\multicolumn{3}{c}{Discriminator Network}\\  \midrule
FF & 43,776 & 156 \\
FF & 65,792  & 256   \\
FF & 257  & 256   \\ \bottomrule
Total & 293,547  & 1
\end{tabular}
\label{table_c10_params_prop}
}
\caption{Parameters per layer for a \acs{GAN}, trained using \acp{AIR} in the corresponding representation.}
\label{table_c10_params}
\end{table}

\begin{figure}
	\centering
	\subfloat[\acp{GAN} trained using \ac{FIR} taps of \acp{AIR}]{
		\includegraphics{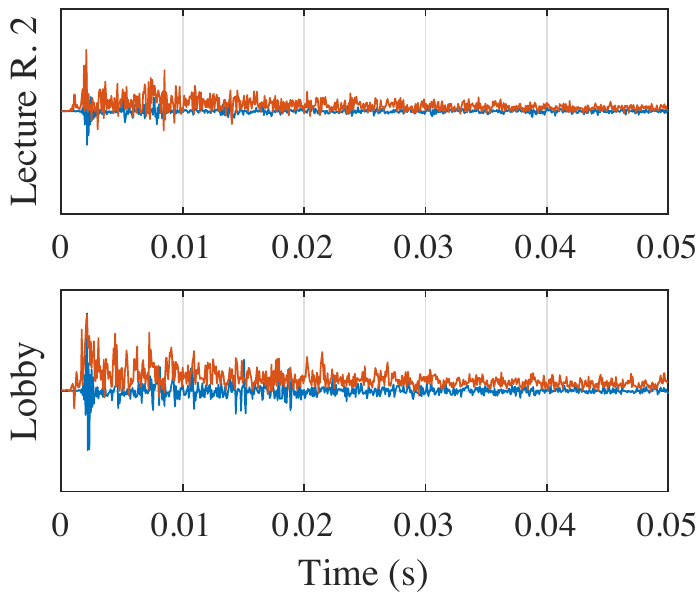}
	\label{fig_c10_e2e_vs_real}
		}\ifIEEE
		
		\else
		~~~
		\fi
\subfloat[\acp{GAN} trained using the proposed low-dimensional representation. The generated \acp{AIR} in this  representation are used to  construct the taps of \ac{FIR} filters, shown in the plots.]{
	\includegraphics{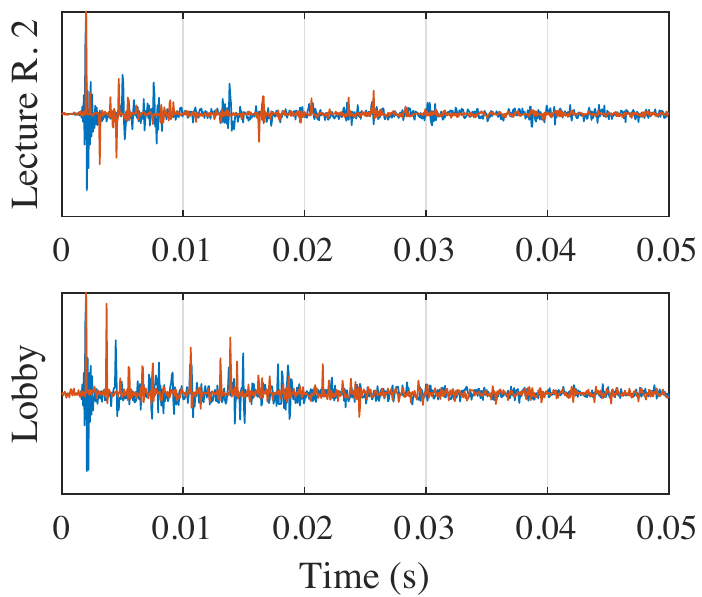}
	\label{fig_c10_prop_vs_real}
	}
	\caption{Comparison between  artificially generated and measured  \acsp{AIR}. Artificial \acp{AIR} are  generated by  \acsp{GAN}, trained using  \acp{AIR} measured in the real rooms. }
		\label{fig_c10_gen_vs_real}
	\end{figure}

In this work, the generation of \acp{AIR} is based on training one \ac{GAN} for each of the 7 rooms, part of the training database. Therefore, 7 \acp{GAN} are trained and each one of them is used to generate a number of \acp{AIR} as if they were measured in the corresponding rooms. The above process is repeated 2 times, with the representation of the \acp{AIR} passed to the \acp{GAN} during training changed between the two. The two representations considered are the raw \ac{FIR} taps and the parameters of the low-dimensional representation proposed in Section \ref{section_low_dim_rep}. The number of parameters composing the discriminator and generator of the \acp{GAN} for each of the two cases is given in Table \ref{table_c10_params}. In Figure \ref{fig_c10_gen_vs_real}, generated artificial responses from \acp{GAN} trained using the two representations are visualised along with responses measured in the real rooms.

\begin{figure}
	\centering
        \subfloat[Epoch 80]{
                \includegraphics[scale=0.65]{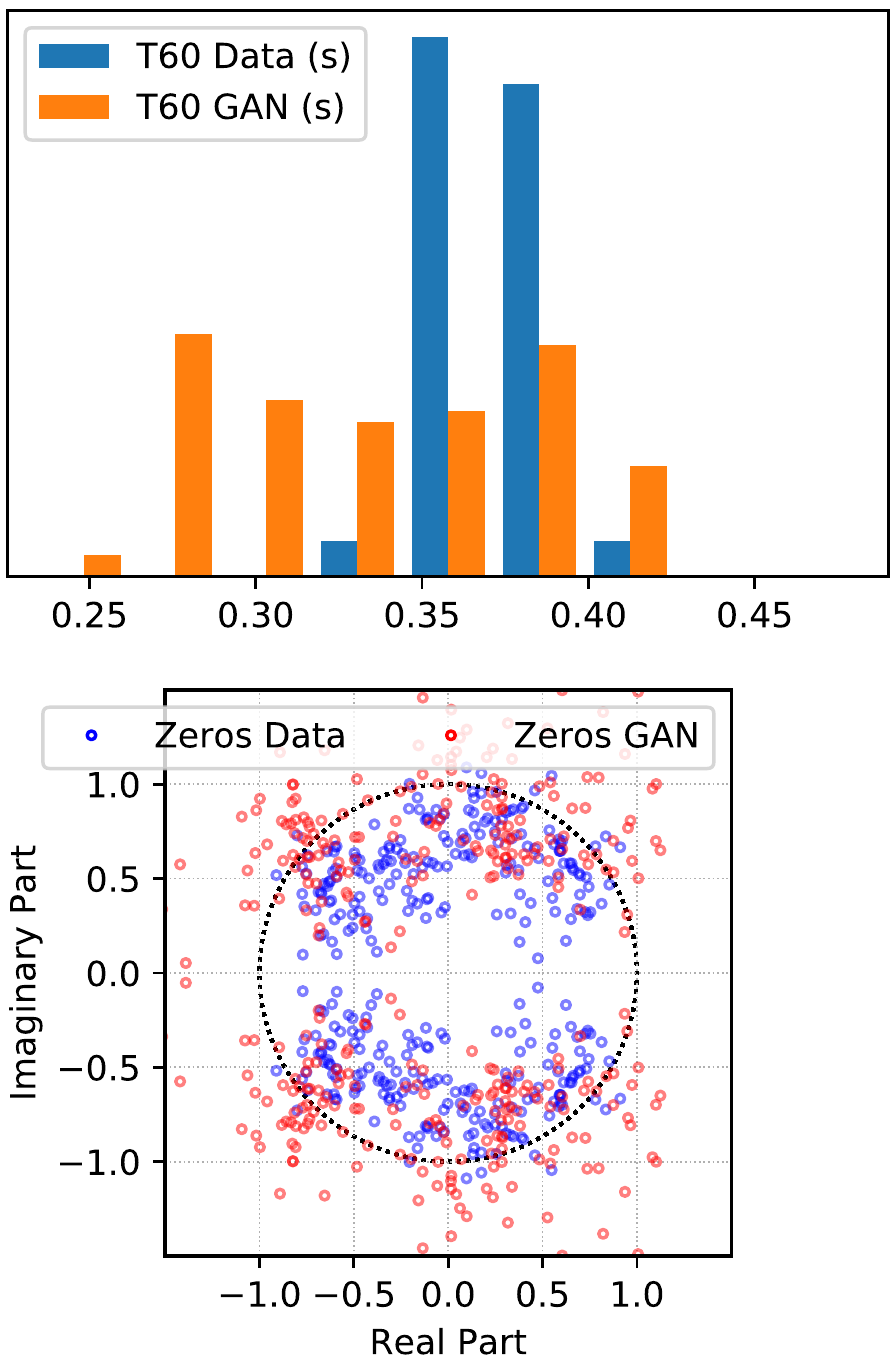}
                \label{figure_c10_gan_evolution_1}
        } \ifIEEE   
        
        \else
        ~
        \fi
                \subfloat[Epoch 960]{
                \includegraphics[scale=0.65]{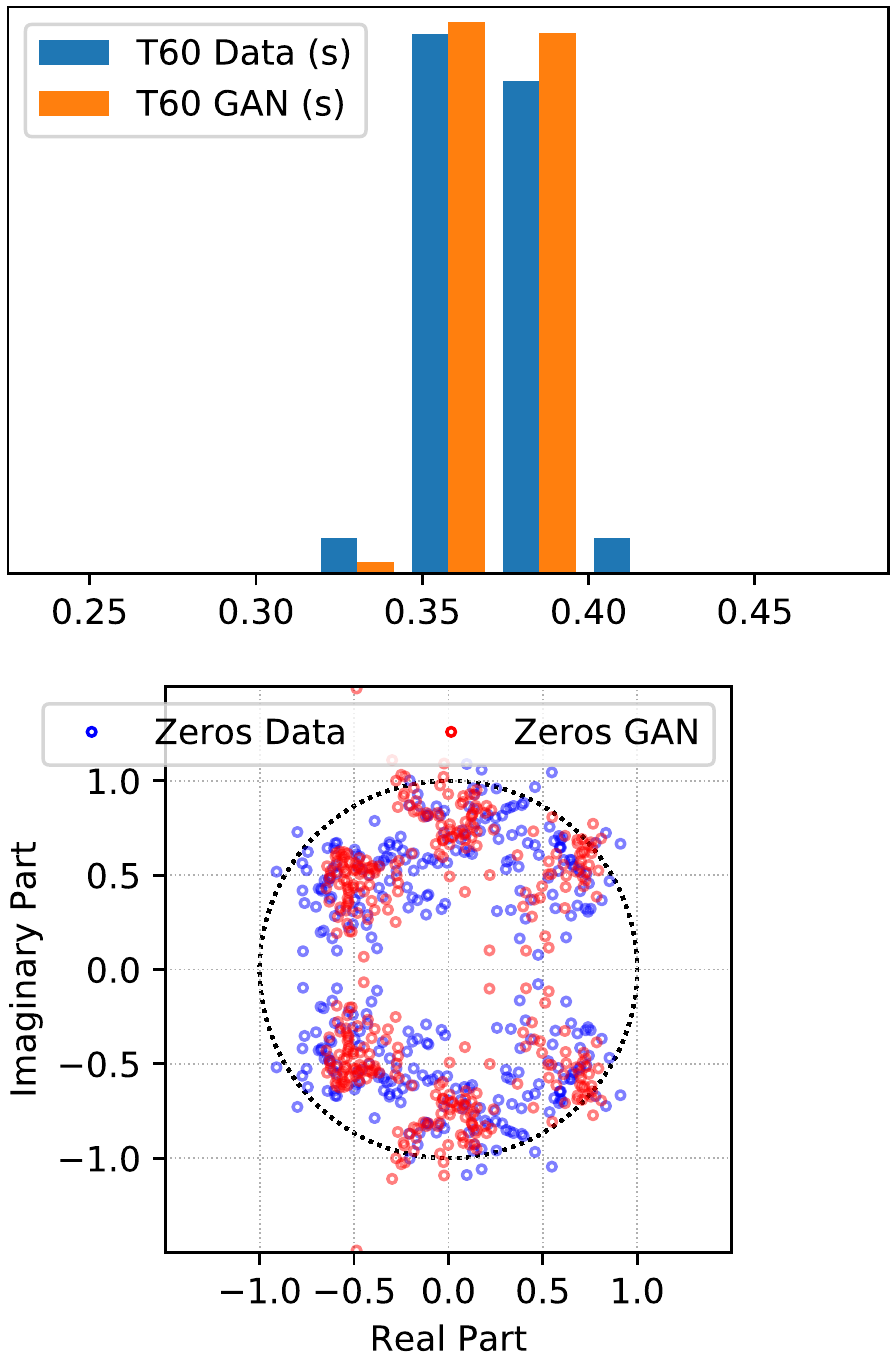}
                \label{figure_c10_gan_evolution_4}
    	}
	\caption{Evolution of the distribution of generated $T_{60}$ values and the frequency of the zeros of the generated  \acs{IIR} filters during the training of  a \acs{GAN}, using the proposed low-dimensional representation of \acp{AIR} from a meeting room.}
	\label{figure_c10_gan_evolution}

\end{figure}

One of the critically discussed issues in the literature with regard to the training of \acp{GAN} is the lack of established methods for evaluating the generated data \cite{Salimans2016}. The data can always be evaluated for their usefulness for data augmentation in terms of the increase in classification accuracy but a way to evaluate the generated samples directly saves unnecessary classifier training times.  Indeed, in this work, it would also be of interest to evaluate how realistic the generated responses are before using them to train a complex room classifier. Evaluating realism is difficult  and it is very hard to quantify or even precisely define. However, using the proposed representation as the space for \acp{GAN}  enables the  inspection of semantically meaningful  properties of the generated environments by humans. For instance, when training a \ac{GAN} to learn how to generate responses from a  meeting room, the visualisations of Figure \ref{figure_c10_gan_evolution} can be directly made. The  Figure shows  how  the zeros of the \ac{IIR} filters and  the $T_{60}$ are distributed for the data generated by the \ac{GAN}. The distributions are shown at two stages during  training, at epoch 80 and then much later at epoch 960.  Observing the plots shows  that the distribution of the parameters starts from a near random state and then becomes very similar to that of the data measured in the real room. This is positive evidence that the distribution of these parameters is realistic.

\label{section_air_gen_discussion}

\begin{figure}[t!]
\centering
\begin{minipage}{0.45\textwidth}
\centering
\vspace{0pt}
                \includegraphics[scale=0.8]{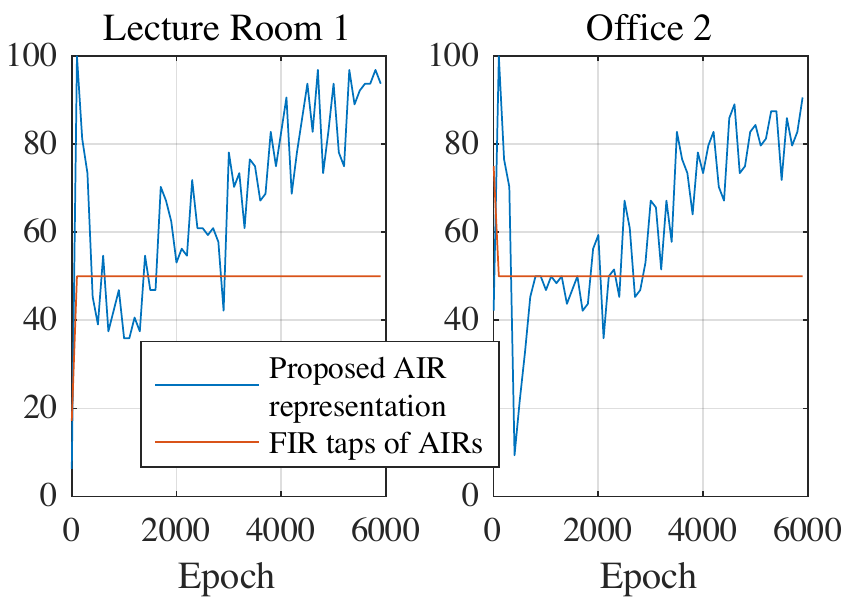} 
        \caption{Comparison of the accuracies of \acs{GAN} discriminators for the case of  training using \acp{AIR} in the proposed low-dimensional representation and for the case of using the taps of the \ac{FIR} filters of \acp{AIR}.}
               \label{fig_c10_prop_vs_e2e_accuracy}

\end{minipage}\ifIEEE

\else
\hfil
\fi
\begin{minipage}{0.48\textwidth}
\centering
\vspace{0pt}
                \includegraphics[scale=0.45]{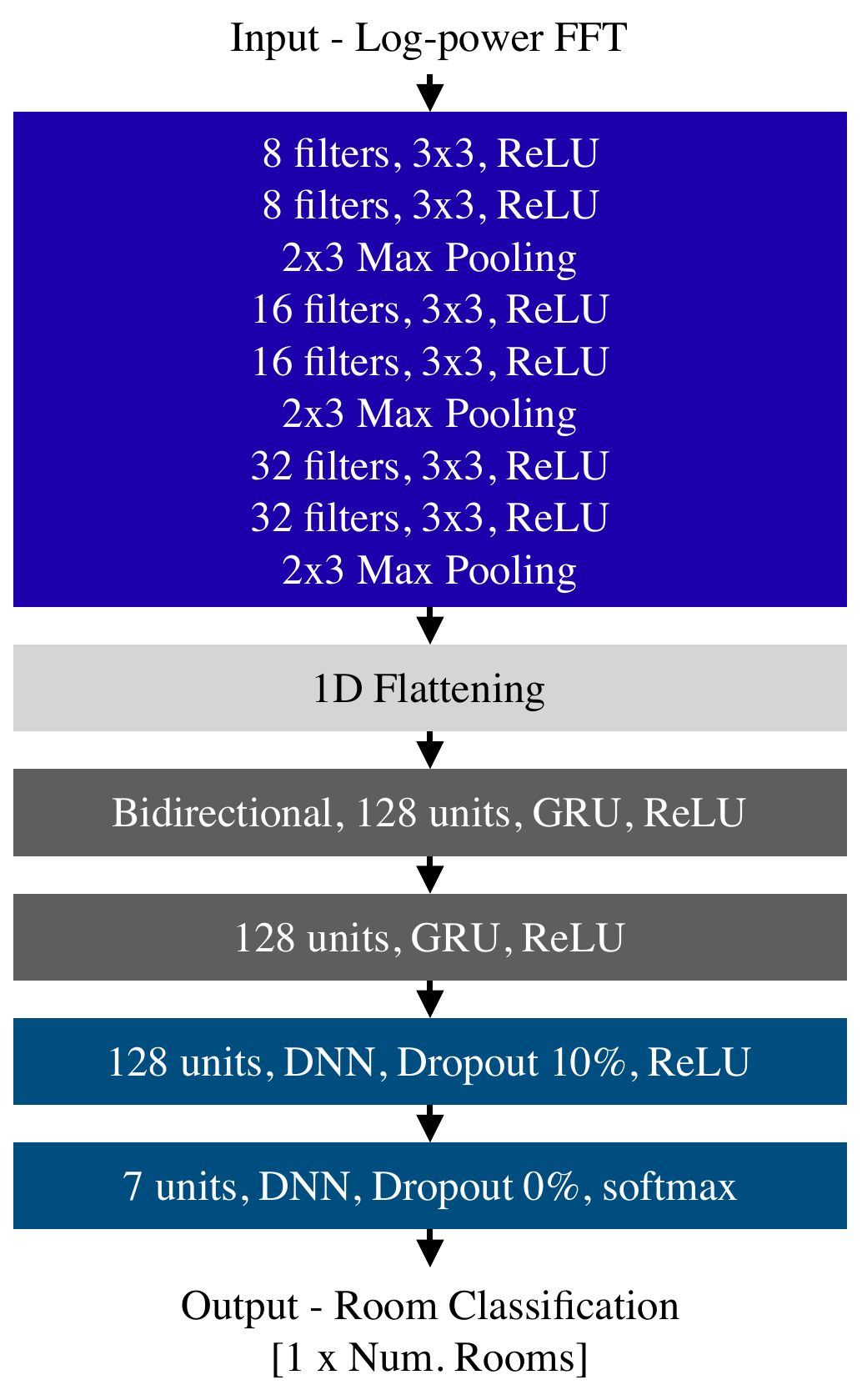} 
        \caption{\acs{CNN}-\acs{RNN} models for room classification from reverberant speech input.}
               \label{figure_c10_models_speech_cnn_rnn}
 \end{minipage}
\end{figure}

Using the raw \ac{FIR} taps of  \acp{AIR} directly to train a \ac{GAN} provides no semantically meaningful  information about the  acoustic environment, in contrast to using the proposed representation. The results of Figure \ref{fig_c10_e2e_vs_real} is the only evidence for the quality of the results and it is inspected directly to analyse them. What can be observed from looking at the real \acp{AIR}, given by the blue lines  in the Figure, is that they are composed of a direct path, sparse reflections in the early part and a decaying envelope. However, the same does not apply for the generated \acp{AIR} that  resort to tracking the overall energy envelope as an approximation to the overall shape. The discriminator of Figure \ref{figure_c10_discriminator} is therefore \textit{fooled} by a simple imitation of the energy envelope of the inputs into believing that they are real \acp{AIR}.  In reality, the generated responses fail  to capture the sparsity of the early part. The opposite is true for the case of training \acp{GAN} using the proposed representation where the sparse nature of the early part is well captured and so is the tail, as shown in Figure \ref{fig_c10_prop_vs_real}.

Probing into the process of training the network reveals important information, which explains the above observations. Figure \ref{fig_c10_prop_vs_e2e_accuracy} shows the accuracy of 4 discriminators that describe the cases of training a  \ac{GAN} for 2 rooms, using each of the 2 representation. The case of training \acp{GAN}  using \ac{FIR} taps shows the discriminator being unable to discriminate between real and fake samples after a small number of epochs. The accuracy plateaus to 50\%, which indicates that the discriminator is making a random decision between  \textit{real} and \textit{fake}. A weak discriminator cannot yield a stronger generator as Nash equilibrium is reached at this point \cite{Goodfellow2014}. The opposite is true for the case of using the proposed representation, as the discriminator's accuracy almost continuously increases and reaches values as high as 90\%. This is attributed to the high dimensionality of the raw \acp{AIR}. This causes the  \ac{GAN} to scale-up to more than 17~million parameters and the small amount of training data does not allow for the training of very large networks that would result from adding more layers. This actually brings this work back to its original motivation, which was high-dimensionality and the lack of large data availability. This reinforces the need for low-dimensional and informative representations for \acp{AIR}, such as the one proposed in this paper. 

The experiments above have discussed how \acp{GAN} are trained to generate \acp{AIR} for a set of rooms. One  \acp{GAN} is trained for each of the 7 rooms, part of the training data. Each network then generates a set of artificial responses as if they were measured in the real room.  The following experiments will show how the generated responses are used as a data augmentation dataset to tackle the small availability of training data in order to improve the accuracy of a \ac{DNN} room classifier.

\subsection{Data augmentation for  \acs{DNN} room classifiers}

Room classification using state-of-the-art classifiers was investigated in \cite{Papayiannis2018a}. \acp{DNN} were used to classify a reverberant speech signal in terms of the room  it was recorded in. The investigation has shown that the limited availability of \acp{AIR} and their high-dimensionality limit the performance of classifiers. This paper proposes novel methods that lead to an increase in the availability of training data in the form of \acp{AIR}, with the aim to increase the accuracy of \ac{DNN} classifiers. The experiments above have shown how \acp{GAN} are used to create artificial \acp{AIR}, given a set of real ones. These \acp{AIR} are used as part of the proposed  data augmentation method.

\subsubsection{Classifier \ac{DNN}}

\ifIEEE
\newcommand{\thisenvb}{figure*}
\else
\newcommand{\thisenvb}{figure}
\fi

\begin{\thisenvb}[t]
\centering
	\includegraphics{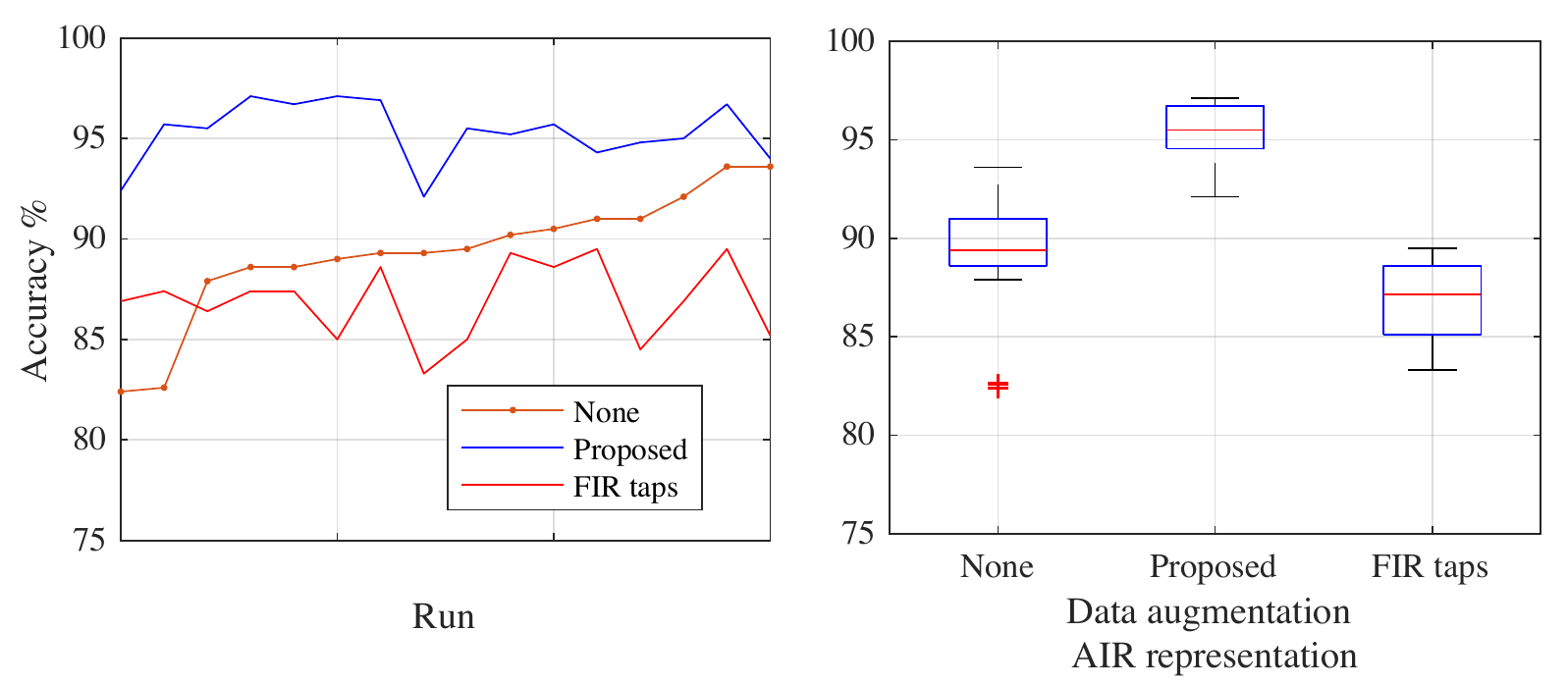}
	\caption{Accuracy of room classification \acsp{DNN} for the cases of using no data augmentation  (baseline), the proposed data augmentation method and data augmentation using the raw \ac{FIR} taps of \acp{AIR}. Top results are sorted from worst to best with regards to the baseline.  Individual runs indicate  training  on a different machine, all with NVIDIA Tesla K80 \acp{GPU}.  }
	\label{fig_c10_accuracy_with_da}
\end{\thisenvb}

The \ac{DNN}   classifier used in this experiment was proposed in \cite{Papayiannis2018a} for the task of room classification. The model is a \acs{CNN}-\acs{RNN} and it is shown in Figure \ref{figure_c10_models_speech_cnn_rnn}.  The training process for the classifier network is as proposed in \cite{Papayiannis2018a}.

 The network is trained and evaluated in 3 configurations. The baseline configuration  involves no data augmentation. It only uses  measured \acp{AIR} from the \ac{ACE} database \cite{Eaton2015d} for training. The second configuration uses data augmentation  done using artificial \acp{AIR} generated by \acp{GAN}  trained using the raw \ac{FIR} taps of \acp{AIR}. Finally, in the last configuration, data augmentation is done using the  method proposed in this paper. Each configuration is evaluated based on its  accuracy on a test set, which is discussed below. The training of classifiers of each of the 3 configurations is repeated 16 times on  16 different machines to average the effect of different initialisations. The training and test data were the same across machines. The hardware used was   NVIDIA Tesla K80 \acp{GPU}.

\subsubsection{Training and test data}

The \ac{ACE} database \acp{AIR}  \cite{Eaton2015d} are used for this experiment. They are segmented to reserve a test set prior to training. The \ac{ACE} database consists of a total of 700 \ac{AIR}, recorded in 7 rooms. The 42 \acp{AIR}, which were recorded using the \textit{Mobile} microphone array are reserved for testing. The remaining 658  \acp{AIR} are all used to train the relevant \acp{GAN} and artificially create further \acp{AIR} for  data augmentation. They contain 94 \acp{AIR} per room. The proposed data augmentation method involves the training of 1 \ac{GAN} for each of the 7 rooms. Each \ac{GAN} is trained using  94 \acp{AIR} and it is used to generate an additional 100 \acp{AIR}. This results in an additional 700 \acp{AIR} in total, doubling the size of the \ac{ACE} database. The training data along with the artificially created \acp{AIR}, form the  training set in the proposed  method.

\ifIEEE
\newcommand{\thisenvc}{figure*}
\else
\newcommand{\thisenvc}{figure}
\fi

\begin{\thisenvc}[t]
	\centering
	\includegraphics{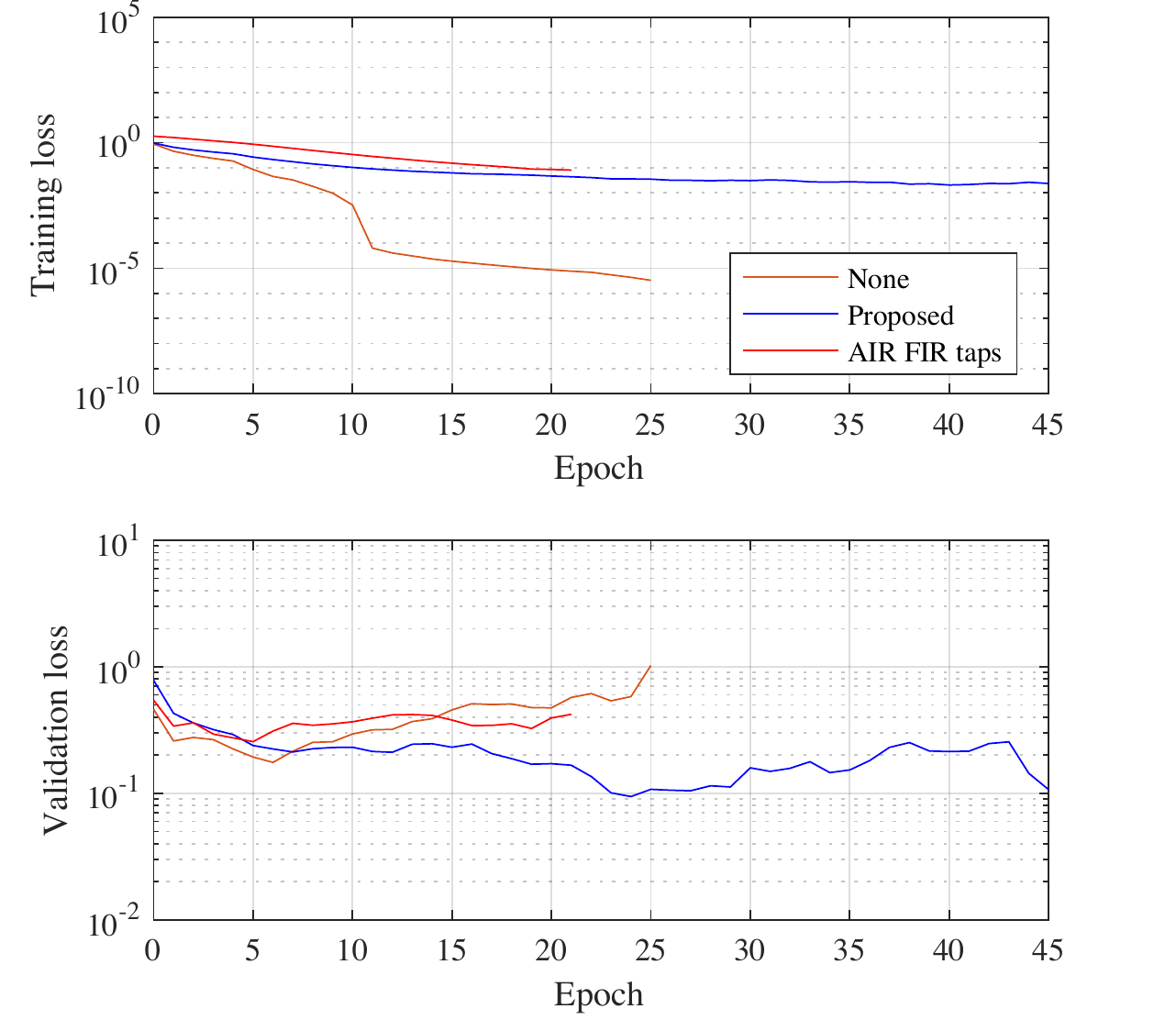}
	\caption{Comparison of the training and validation losses for a room classification \acs{DNN} for the cases of using no data augmentation, the proposed data augmentation and the data augmentation using \ac{FIR} taps of \acp{AIR}. Losses are smoothed using a moving average  window of 10 epochs. }
	\label{fig_c10_losses_with_da}
\end{\thisenvc}

The experiment is investigating the classification of reverberant speech in terms of the room where the recording took place.  All training \acp{AIR} are therefore convolved with 20  speech utterances each of length 5~s, taken  from the \acs{TIMIT} database. This is identical to the process of \cite{Papayiannis2018a}. Train and test speech and speaker databases are separated and are not mixed. Each utterance contains only one speaker.  Convolving new speech samples with the data augmentation \acp{AIR} will introduce an additional variable in the comparison of the results. Therefore, to avoid this, the same exact speech utterances  convolved with the measured \acp{AIR} for each room are  convolved with the data augmentation \acp{AIR} for the corresponding rooms. The 42 test \acp{AIR} are convolved with 10  utterances each, again of length 5~s. The test and train reverberant speech are consistent throughout all the experiments and the only variable is the addition of the data augmentation \acp{AIR}. All data is sampled at 16~kHz.

The segmentation of the available data into test and training sets is discussed above. Neither the speaker, the speaker's  position,   or the microphone array used to construct the test data, were presented to the classifier during training.  The artificial \acp{AIR}, generated by the \acp{GAN},  were generated as if they were measured in  the same rooms as the training data but at different positions. The addition of these \acp{AIR} generated by the  \acp{GAN}  aims to improve the classification test accuracy. Data augmentation performed in this fashion provides class invariant transformations of the training data to the classifier.  Therefore, this experiment evaluates the improvement in the generalisation of the classifier offered by the proposed data augmentation method.

\subsubsection{Results}

The results of  evaluating the proposed method are shown in Figure \ref{fig_c10_accuracy_with_da}, in terms of the classification accuracy on the test set. The results show that the  proposed method outperforms the baseline in all runs. The median accuracy of the baseline is 89.4\%, the proposed method's is 95.5\% and the \ac{AIR} tap based method's is 87.15\%. Therefore, the proposed method increases the accuracy of the room classifier. The increased accuracy is not attributed to an increase of the speech data as the exact same speech samples were used in all 3 cases. The use of the high-dimensional raw \ac{AIR} taps proved even less effective than the baseline and the trained \acp{GAN} involved a total of around 17~million parameters. The proposed domain increased the classification accuracy while using  only  0.29~million parameters.  

 To understand how the proposed data augmentation  method affects the training, the training and validation losses of the trained \acp{DNN} are  shown in Figure \ref{fig_c10_losses_with_da}. The baseline training loss, which uses no data augmentation,  shows that after 10 epochs the model starts to overfit. The training loss starts to substantially decrease and the validation loss increases.   However, when using the proposed data augmentation, the validation loss continues to follow a decreasing trend for longer  and the training loss is approximately monotonically decreasing.  Therefore, the data augmentation method improves generalisation by providing a meaningful and realistic interpolation of the available \acp{AIR} in a low-dimensional manifold of the reverberation effect.

The presentation of the results of the experiments concludes this Section. The final Section will review the contributions of this paper and provide a conclusion.

\section{Conclusion}
\label{chapter15}

This paper has proposed a novel method for data augmentation for the training of \ac{DNN}   room classifiers. The proposed method relies on the training of \acp{GAN}, using \acp{AIR} in a proposed low-dimensional representation. The representation combines parameters of the early reflections  and established parameters for late reverberation. The \acp{GAN} are used to create artificial \acp{AIR} from a set of known rooms.    The proposed method enabled \acp{GAN} to generate  artificial responses with realistic features, able to capture the sparse  properties of the early reflections and the decaying tail.  In the experiments presented, the proposed method  increased the accuracy of a \acs{CNN}-\acs{RNN} room classifier from 89.4\% to 95.5\%, when compared to the case of using no data augmentation. 

The training of \acp{GAN} as proposed in this work uses \acp{AIR} measured   in rooms in order to create a number of artificial but realistic \acp{AIR}. This process finds applications beyond room classification. Artificial reverberation applications \cite{Valimaki2012} can benefit from such approaches, where a number of artificial environments with specific properties can be created by training \acp{GAN} using a specific modality of acoustic environments. For instance, providing a \ac{GAN} with enough \acp{AIR} from many concert halls will enable it to learn to generate many more artificial \acp{AIR} from  many artificial concert halls. The possibilities for such  methods are numerous.

\bibliographystyle{ieeetr}
\bibliography{ms}

\end{document}